\documentclass[nofootinbib,onecolumn,aps]{revtex4}
\usepackage{amsmath}
\usepackage{graphicx}
\usepackage{amssymb}
\usepackage{float}
\usepackage{mathpazo}
\usepackage{lineno}

\pdfoutput=1
\newcommand{\ignore}[1]{}

\newcommand{\be}{\begin{equation}}
\newcommand{\ee}{\end{equation}}
\def \ba#1\ea{\begin{align}#1\end{align}}
\newcommand{\bit}{\begin{itemize}}
\newcommand{\eit}{\end{itemize}}

\def \slashb#1{\setbox0=\hbox{$#1$}#1\hskip-\wd0\dimen0=5pt\advance
        \dimen0 by-\ht0\advance \dimen0 by\dp0\lower0.5\dimen0\hbox
          to\wd0{\hss \sl/\/ \hss}}

\input{epsf}
\input{tcilatex}
\begin{document}

\title{$\sin ^{2}\theta _{W}$ \textbf{estimate and neutrino electromagnetic
properties from low-energy solar data}}
\author{Amir N. Khan}
\email{ntrnphysics@gmail.com}
\email{akhan@fnal.gov}
\affiliation{Max-Planck-Institut fu%
\"{}%
r Kernphysik, Saupfercheckweg 1, 69117 Heidelberg, Germany}
\affiliation{Theoretical Physics Department, Fermi National Accelerator Laboratory, P.O.
Box 500, Batavia, IL 60510, USA}

\begin{abstract}
We report new values of weak-mixing angle ($\sin ^{2}\theta _{W}$)$,$
neutrino effective magnetic moment ($\mu _{\nu }^{eff}$) and the charge
radii $(\left \langle r_{\nu _{\alpha }}^{2}\right \rangle )$ using the
lowest-energy (to-date) solar neutrino data of$\ $pp, $^{7}$Be and pep
spectra from phase-I and phase-II runs of Borexino experiment. The best-fit
values obtained are $\sin ^{2}\theta _{W}=$0.235$\pm $0.019 at 1$\sigma \ $%
with a precision comparable to that of the combined reactor and accelerator
very short-baseline experiments and $\mu _{\nu }^{eff}\leq 8.7\times
10^{-12}\mu _{B}$ \ at 90\% C.L. with a factor of 3 improvement over the
previous bounds. This leads to the improvement of constraints on all the
related magnetic moment matrix elements for the Majorana-type and Dirac-type
neutrinos in mass basis and also stronger bounds on the\ magnetic moment
flavor states. The bounds on the neutrino charge radii obtained are $%
0.82\times 10^{-32}$cm$^{2}\leq \left \langle r_{\nu _{e}}^{2}\right \rangle \
\leq 1.27\times 10^{-32}$ cm$^{2}\ $and\ $-9\times 10^{-32}$cm$^{2}\leq
\left \langle r_{\nu _{\mu },\nu _{\tau }}^{2}\right \rangle \  \leq 3.1\times
10^{-31}$ cm$^{2}\ $at 90\% C.L..
\end{abstract}

\date{\today }
\pacs{xxxxx}
\maketitle

\section{Introduction}

Tests of the Standard Model (SM) of particle physics, and the corresponding
determination of its characteristic electroweak parameters, span many orders
of magnitude in energy. The measurements of the scale dependence of the
fundamental electroweak parameter $\sin ^{2}\theta _{W}$\ now cover the
range from 10 TeV at the Large Hadron Collider CMS \cite{cms}, ATLAS \cite%
{atlas}, LHCb \cite{lhcb0} to a few MeV and below at Solar neutrino
detectors SNO \cite{SNO}, KamLAND \cite{Kaml}, Borexino \cite%
{brxpp,brxbe,brxpe,brxII} and atomic spectral measurements of electroweak
parity violation in $^{133}$Cs \cite{Cs,boulder,porsev}. In between, in the
range from multi-MeV to multi- GeV is a host of experiments in operation or
development designed to look for deviations from the SM in the form of
lepton flavor and universality violations \cite{lhcb1,lhcb2,clfv} and to
make precision measurements in the process (Belle, BelleII, BaBar)\cite%
{lhcb3}, Q-weak \cite{QweakI,QweakII}, SoLID \cite{SoLID}, MOLLER \cite%
{Moller}. At present the lowest energy determination of $\sin ^{2}\theta
_{W} $\ is that provided by the parity-violation measurement in $^{133}$Cs
at 2.4 MeV \cite{Cs,boulder,porsev} and in the lowest possible energy solar
data at 1.4MeV and below has been estimated in ref. \cite{ANKS}.

In the low energy regime ($\leqslant 100$MeV), the neutrino interactions in
scattering off electrons have played a key role in understanding the gauge
structure, precision test of the standard model and in looking for new
physics like the electromagnetic properties and nonstandard interactions of
neutrino \cite{ANKS,amirgl}. One the other hand, the neutrino oscillation
experiments have also entered into the precision era of the oscillation
parameters \cite{JUNO,HKexpt,CJPL}. As the oscillation parameters get more
and more precise, the experimental outputs give as a by-product several
other interesting precision tests of the SM parameters including $\sin
^{2}\theta _{W}$\ and increase the room new physics such as electromagnetic
properties and the nonstandard interactions of neutrinos as recently been
explored in ref. \cite{ANKS} for Phase-II data of Borexino's measurements.

The low-energy neutrinos scattering off the electrons elastically are ideal
sources for measuring the value of $\sin ^{2}\theta _{W}$\ and for exploring
the electromagnetic properties of neutrinos \cite{pvogel}. Here we focus on
low energy neutrino-electron scattering as a probe for new features in
electroweak physics beyond the standard model. SM loop calculations predict
that the neutrinos have finite, but exceedingly small, magnetic moment and
charge radius. At tree level, their only interaction with the electron is
purely weak and short range, while the higher order electromagnetic cross
section is long range, inversely proportional to the electron recoil energy.
This kinematic feature enhances the photon exchange in low energy
neutrino-electron scattering compared to exchange of the weak bosons and
makes neutrino-electron scattering an ideal, sensitive testing ground for
this new physics, which shows up as anomalous distortions of the electron
recoil spectrum. It is this feature that we pursue with the pp (0-0.480MeV), 
$^{7}$Be(0.86MeV) and\ pep (1.4MeV) data at the very low end of the solar
neutrino spectrum, detected by Borexino's clean electron recoil detection
system in phase I \cite{brxpp,brxbe,brxpe} and phase II \cite{brxII} runs
and use it for the precision test of $\sin ^{2}\theta _{W}$\ and for the
study of electromagnetic properties.

In the next section, we review the formalism for $\nu _{\alpha }e-$\
electroweak scattering process in the presence of electromagnetic diagram
and in sections III and IV we discuss the solar neutrino oscillation
probabilities on Earth and the neutrino magnetic moments in the flavor and
mass basis both for Dirac and Majorana type neutrinos, respectively. Section
V is dedicated to the analysis details and to the obtained results while we
conclude this study in section VI.

\textit{\ }

\begin{figure}[t]
\begin{center}
\includegraphics[width=3.2in]{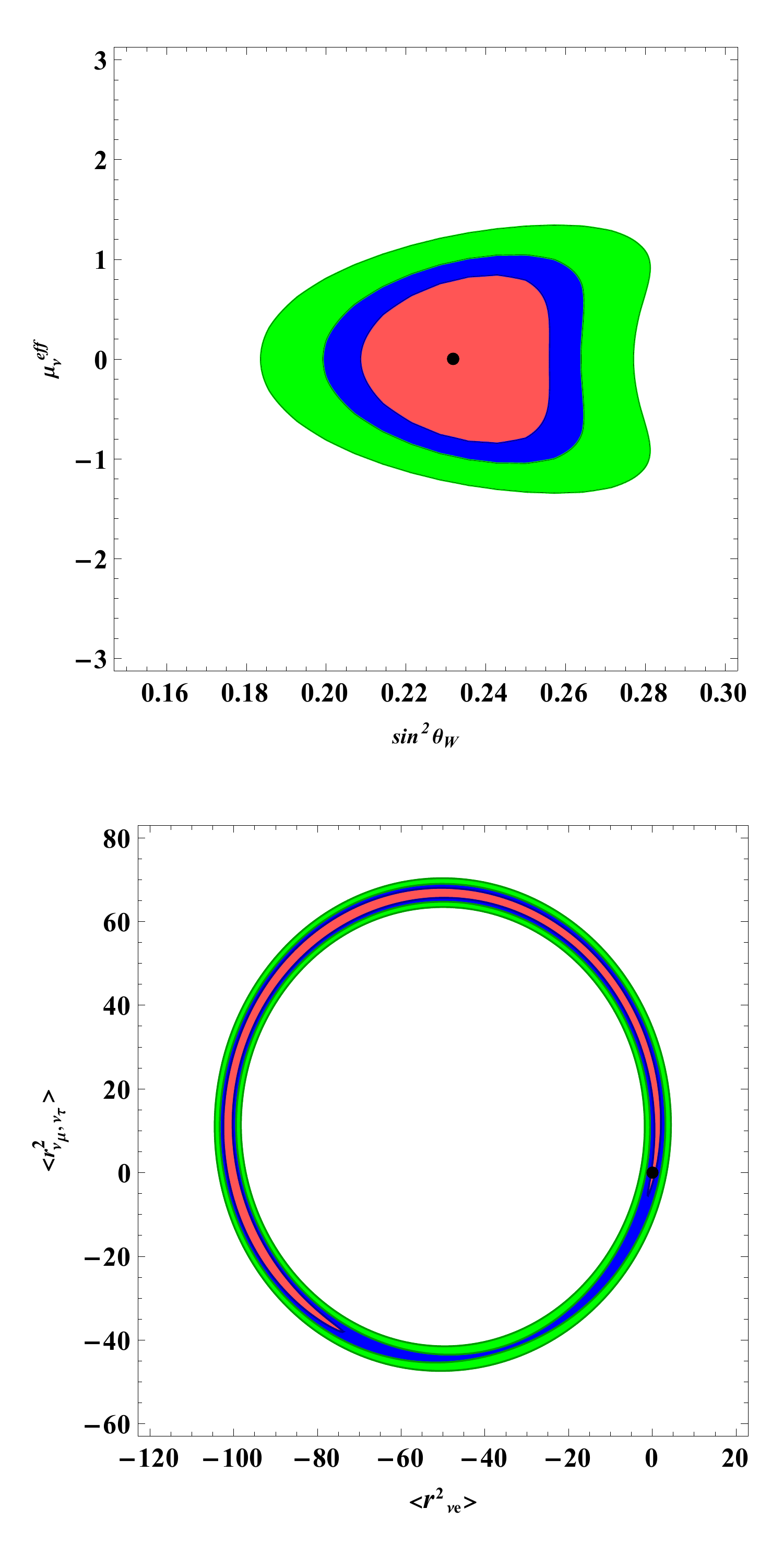}
\end{center}
\caption{\textbf{\ }The 2-\textit{d.o.f} parameter space of weak-mixing
angle and effective magnetic moment (upper) and charge radius of $\protect%
\nu _{e}$ versus the charge radius of $\protect \nu _{\protect \mu }$ or $%
\protect \nu _{\protect \tau }$ (lower) using the solar low-energy data of 
\textit{pp, }$^{7}$\textit{Be, pep} reactions at Borexino phase-I and
phase-II runs. The red, blue, green regions are 1$\protect \sigma $ and 90\%,
95\% C.L. boundries. The black dot shows the best-fit points. The scale of $%
\protect \mu _{\protect \nu }^{eff}$ is in units of $\times $10$^{-11}\protect%
\mu _{B}$ and for the charge radii is in $\times $10$^{-32}$cm$^{2}.$}
\end{figure}
\ 

\section{$\protect \nu _{\protect \alpha }e-$\ scattering cross-sections%
\textit{\ }}

The differential cross-sections for the three processes of $\nu _{e}e-,\  \nu
_{\mu }e-$\ and $\nu _{\tau }e-$\ scatterings are the incoherent sums of the
standard model weak and the electromagnetic interaction differential
cross-sections. After integration over the recoiled electron energy ($T$),
the total cross-sections read \cite{pvogel}

\bigskip 
\begin{equation}
\sigma _{tot}=\int_{0}^{T_{max}}\left[ (\frac{d\sigma _{\nu e}}{dT})_{SM}+(%
\frac{d\sigma _{\nu e}}{dT})_{_{em}}\right] dT,
\end{equation}%
where

\begin{equation}
(\frac{d\sigma _{\nu e}}{dT})_{SM}=\frac{2G_{F}^{2}m_{e}}{\pi }%
[g_{L}^{2}+g_{R}^{2}\left( 1-\frac{T}{E_{\nu }}\right) ^{2}-g_{L}g_{R}\frac{%
m_{e}T}{E_{\nu }^{2}}]
\end{equation}%
and%
\begin{equation}
(\frac{d\sigma _{\nu e}}{dT})_{_{em}}=\frac{\pi \alpha _{em}^{2}(\mu _{\nu
}^{eff})^{2}}{m_{e}^{2}\mu _{B}^{2}}[\frac{1}{T}-\frac{1}{E_{\nu }}].
\end{equation}%
Here 
\begin{equation*}
g_{L}=(g_{V}+g_{A})/2+1\text{,\  \ }g_{R}=(g_{V}-g_{A})/2
\end{equation*}%
for $\nu _{e}$ and%
\begin{equation*}
g_{L}=(g_{V}+g_{A})/2\text{,\  \ }g_{R}=(g_{V}-g_{A})/2
\end{equation*}%
for $\nu _{\mu }$ and $\nu _{\tau },$ where $g_{V}=-1/2+2\sin ^{2}\theta
_{W} $ and $g_{A}=-1/2$, $\mu _{\nu }^{eff}$ is the effective neutrino
magnetic moment, $\mu _{B}$ is the Bohr magneton unit, $\alpha _{em}$ is the
fine-structure constant, $m_{e}$ is the electron mass, $E_{\nu }$ is the
neutrino energy and $T^{\max }$ is the maximum recoiled-electron energy in
the detector.$\ T^{\max }(E_{\nu })\equiv E_{\nu }/(1+m_{e}/2E_{\nu })$,
where $0<E_{\nu }<0.420$ MeV for $pp$ events and $E_{\nu }=0.862$ MeV and $%
1.44$ MeV for $^{7}$Be and $pep$ events, respectively. Notice that the term $%
"1"$\ in the defintion of $g_{L}$\ for $\nu _{e}$\ corresponds to the CC\
contribution in the $\nu _{e}-e$\ scattering.

Neutrinos are electrically neutral particles in the standard model, their
electric form factors in terms of neutrino charge radius can still give
useful information about the electromagnetic properties. In the earlier
studies, it was claimed that neutrino charge radius is not a physical
quantity because of the ultraviolet-divergences produced in the unitary
gauge \cite{bard} or in general gauge \cite{DvoI, DvoII} in one-loop and $%
\gamma -Z$\ self-energy calculations. These infinities can be cured using
the unitary gauge if the neutrino-lepton neutral current contribution is
added to usual terms, then for the neutrino charge radius a finite
gauge-dependent quantity can be obtained \cite{leeI}. In order to calculate
the neutrino charge radius as a finite gauge-independent physical quantity,
one has to add box diagram contribution to the tree level diagrams of the
neutrino-lepton scattering processes\textbf{\ }\cite{leeII}. The neutrino
charge radius can thus be introduced as gauge-independent physical
observable in the calculations of one-loop approximation including the
additional terms from the $\gamma -Z$\ boson mixing and the box diagrams of
W and Z bosons \cite{jbI, jbII, jbIII} as 
\begin{equation}
\left \langle r_{\nu _{\alpha }}^{2}\right \rangle =\frac{G_{F}}{4\sqrt{2}%
\pi ^{2}}[3-2\log (\frac{m_{\alpha }^{2}}{m_{W}^{2}})]
\end{equation}%
where m$_{W}$\ is the W-boson mass and m$_{\alpha }(\alpha =e,\mu ,\tau )$\
denotes the masses of charged leptons and $G_{F}$ is the Fermi constant%
\textbf{.}

In order to see the effects of the neutrino charge radius, one can modify
the definition of $g_{V\text{ }}$as $g_{V}=\sin ^{2}\theta _{w}-1/2$\ + $(2%
\sqrt{2}\alpha _{em}/3G_{F})\left \langle r_{\nu _{e}}^{2}\right \rangle $\
for $\nu _{e}-e$\ scattering and $g_{V}=\sin ^{2}\theta _{w}-1/2$\ + $(2%
\sqrt{2}\alpha _{em}/3G_{F})\left \langle r_{\nu \mu ,\nu \tau
}^{2}\right
\rangle ~$\ for $\nu _{\mu }e-$\ or $\nu _{\tau }e-$\
scatterings \cite{pvogel} as was done in ref. \cite{branco} for the reactor
data.

\section{Solar neutrino oscillation probabilities at Earth}

For low-energy solar neutrino spectra of $pp,^{7}Be$ and $pep$, the LMA-MSW
expectation is that the mixing at Earth is essentially due to the vacuum
oscillations. In this case the oscillation amplitude takes the matrix form $%
A_{\alpha \beta }=U_{\alpha a}X_{a}U_{a\beta }^{\dagger }$, where $a,b,c..,$
are the mass basis indices and $\alpha $ and $\beta $ are the flavor basis
indices (summation over repeated indices is implied). The $U$ matrix is the
neutrino mass mixing matrix for any number of neutrinos and $X$ is the
diagonal phase matrix $X\ $= diag$(1,\exp (-i2\pi L/L_{21}^{osc}),\exp
(-i2\pi L/L_{31}^{osc},...)$, where the oscillation length is defined as $%
L_{ab}^{osc}=4\pi E/(m_{a}^{2}-m_{b}^{2})$. The oscillation probability
reads as

\begin{equation}
P_{\alpha \beta }=|A_{\alpha \beta }|^{2}=|U_{\alpha a}X_{a}U_{a\beta
}^{\ast }|^{2},
\end{equation}%
so the average over an oscillation length is then

\begin{equation}
\langle P\rangle _{\alpha \beta }=U_{\alpha a}U_{\beta a}^{\ast }U_{\alpha
a}^{\ast }U_{a\beta }=|U_{\alpha a}|^{2}|U_{\beta a}|^{2},  \label{eq:Pbar}
\end{equation}%
for the average over one cycle of the probability function. For instance,
for the $3\times 3$ mixing in vacuum, the electron survival averaged
probability is $\langle P\rangle _{ee}$ = $s_{13}^{4}+(c_{12}c_{13})^{4}$ + $%
(s_{12}c_{13})^{4}$, where $s_{ij}\equiv \sin \theta _{ij}$ and $%
c_{ij}\equiv \cos \theta _{ij}$.

For these lowest energy solar neutrinos of \textit{pp}$\ $reaction with
energy $E_{\nu }\leq $ 0.420 MeV, the matter effects on the probability $%
(P_{ee})$ that a $\nu _{e}$ survives\ as $\nu _{e}$ in the trip from the
core of the Sun to the detector are negligible, less than a percent
different from the path-averaged over the oscillation length, thus they are
pure vacuum-mixing predictions. However, for the somewhat higher energy line
spectra of $^{7}$\textit{Be }(0.862 MeV) and \textit{pep} (1.44 MeV)
neutrinos, the matter effects are still small, upto 4-5\%, but not entirely
negligible. Therefore, we include the small modifications due to matter
effects\textbf{\ }to the purely vacuum value of $\langle P_{ee}\rangle $.
For this purpose in the $3\times 3$ scenario, the LMA-Wolfenstein matter
effects modify the vacuum oscillation probabilities of $^{7}$Be and pep
neutrinos as

\begin{equation}
\langle P^{m}\rangle _{ee}=s_{13}^{4}+\frac{1}{2}c_{13}^{4}{}(1+\cos 2\theta
_{12}^{m}\cos 2\theta _{12}),
\end{equation}%
where%
\begin{equation}
\cos 2\theta _{12}^{m}=\frac{1-N_{e}/N_{e}^{res}}{\sqrt{%
(1-N_{e}/N_{e}^{res})^{2}+\tan ^{2}2\theta _{12}}}
\end{equation}%
is the effective mixing angle inside the sun, $N_{e}$ is the electron number
density at the center of the sun, $N_{e}^{res}=\Delta m_{12}^{2}\cos 2\theta
_{12}/2E_{\nu }\sqrt{2}G_{F}$\ is the electron density in the resonance
region and $\Delta m_{12}^{2}$\ is the solar mass-squared difference, $%
\theta _{12}$\ is the solar mixing angle. For $pp$ spectrum, we use electron
density at average \textit{pp} neutrino energy and production point in the
above expressions \cite{lopes} and then assuming an exponential decrease in
the density outward from the core in the analytic approximations as
discussed in detail in ref. \cite{lopes} . It is an excellent approximation
for r $>$0.1R$_{solar}\ $\cite{bks}. 
\begin{figure}[t]
\begin{center}
\includegraphics[width=3.2in]{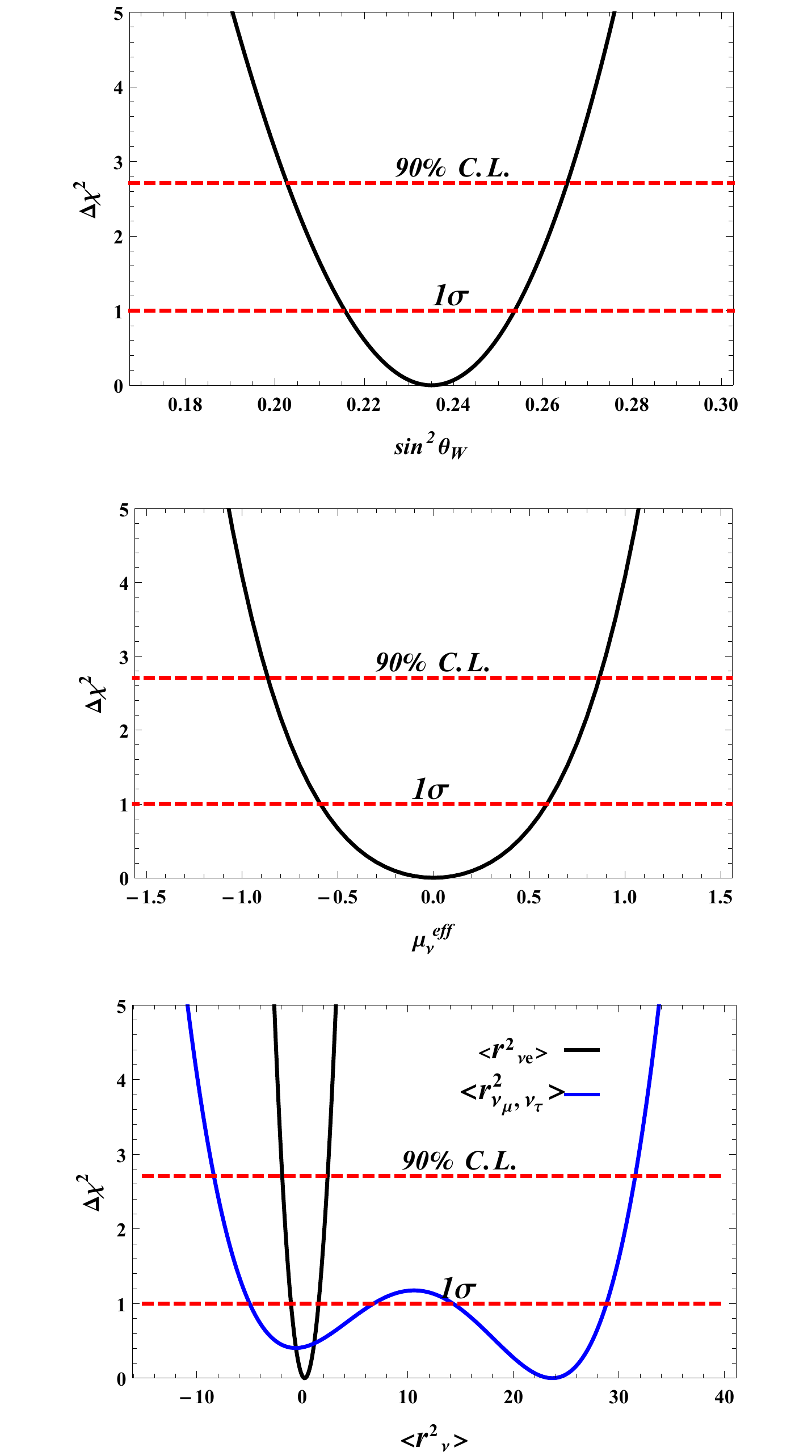}
\end{center}
\caption{{}\textbf{\ }The 1 parameters $\Delta \protect \chi ^{2}$%
-distributions of weak-mixing angle (top), effective magnetic moment
(middle) and the neutrino charge radii of $\protect \nu _{e}$, $\protect \nu _{%
\protect \mu }$ or $\protect \nu _{\protect \tau }$ (bottom) using the solar
low-energy data of \textit{pp, }$^{7}$\textit{Be, pep} reactions at Borexino
phase I and II runs. From bottom to top, the 1$\protect \sigma $ and 90\%
C.L. projections are shown by red dashed lines. The scale of $\protect \mu _{%
\protect \nu }^{eff}$ is in units of $\times $10$^{-11}\protect \mu _{B}$ and
that of $\left \langle r_{\protect \nu }^{2}\right \rangle $ is in $\times $10%
$^{-32}$cm$^{2}.$}
\end{figure}

\begin{table*}[t]
\begin{center}
\begin{tabular}{c|c|c|c|c}
\hline \hline
Parameter & $\sin ^{2}\theta _{W}$ & $\mu _{\nu }^{eff}$ & $\left \langle
r_{\nu _{e}}^{2}\right \rangle $ & $\left \langle r_{\nu _{\mu },\nu _{\tau
}}^{2}\right \rangle \ $ \\ \hline
This work & $0.235\pm 0.019$ & $\leq 8.7\times 10^{-12}\mu _{B}$ & $\  \
[-0.82,\ 1.27$ $]\  \times 10^{-32}$cm$^{2}$ & $\  \ [$\ $-9,31]\times
10^{-32} $cm$^{2}$ \\ \hline
Prediction & $0.23867\pm 0.00016\ $\cite{PDG16} & $\leq 10^{-18}\mu _{B}$ 
\cite{carlo} & $4.1\times 10^{-33}$cm$^{2}\ $\cite{carlo} & $(2.4,1.5)\times
10^{-33}$cm$^{2}\ $\cite{carlo} \\ \hline \hline
\end{tabular}%
\end{center}
\caption{Best-fit value of weak mixing angle$\ $with 1$\protect \sigma $
uncertainty, upper bound on neutrino effective magnetic moment and bounds on
charge radii of $\protect \nu _{e}$ and $\protect \nu _{\protect \mu }$ or $%
\protect \nu _{\protect \tau }$ at 90\% C.L. obtained from the one parameter
at-a-time $\Delta \protect \chi ^{2}$ distributions of Fig. 2. The $3^{rd}$
row shows predicted values. The neutrino effective magnetic moment value is
calculated in the minimally-extended standard model with massive Dirac
neutrinos using the current bound on the neutrino mass. The predicted value
for $\sin ^{2}\protect \theta _{W}$ is the $\overline{MS}$ running parameter
value taken from PDG2016 \protect \cite{PDG16}.}
\end{table*}
\ 

\section{Neutrino magnetic moments in mass and flavor basis}

In the mass-basis, for the Dirac-type neutrinos the effective magnetic
moment have only non-zero diagonal matrix elements while for Majorana-type
of neutrinos, the off-diagonal matrix elements are only relevant \cite%
{boris,shork,valle1,carlo}. Thus, the neutrinos effective magnetic moment
can be written in terms of the Dirac- and Majorana-type magnetic moment
matrix elements in the mass-basis as%
\begin{eqnarray}
(\mu _{eff}^{D})^{2} &=&P_{e1}^{3\nu }\mu _{11}^{2}+P_{e2}^{3\nu }\mu
_{22}^{2}+P_{e3}^{3\nu }\mu _{33}^{2} \\
(\mu _{eff}^{M})^{2} &=&(P_{e1}^{3\nu }+P_{e2}^{3\nu })\mu
_{12}^{2}+(1-P_{e2}^{3\nu })\mu _{13}^{2}+(1-P_{e1}^{3\nu })\mu _{23}^{2}.
\end{eqnarray}%
Notice that we have obtained this form of eq. (9) and (10) after applying
the unitarity condition,$\overset{3}{\underset{i=1}{\  \sum }}P_{ei}=1,$and
assuming the CPT invariance \textbf{(}$\mu _{ij}=\mu _{ji}$\textbf{)}, where 
$P_{e1}^{3\nu }=c_{13}^{2}P_{e1}^{2\nu },P_{e2}^{3\nu
}=c_{13}^{2}P_{e2}^{2\nu },P_{e3}^{3\nu }=s_{13}^{2}$, $P_{e1}^{2\nu }$ and $%
P_{e2}^{2\nu }$ are the effective 2-neutrino solar oscillation probabilities
in the LMA-MSW solution \cite{grim}. Eq. (9) and (10) can be rewritten as

\begin{eqnarray}
(\mu _{eff}^{D})^{2} &=&c_{13}^{2}P_{e1}^{2\nu }\mu
_{11}^{2}+c_{13}^{2}P_{e2}^{2\nu }\mu _{22}^{2}+s_{13}^{2}\mu _{33}^{2} \\
(\mu _{eff}^{M})^{2} &=&c_{13}^{2}\mu _{12}^{2}+(1-c_{13}^{2}P_{e2}^{2\nu
})\mu _{13}^{2}+(1-c_{13}^{2}P_{e1}^{2\nu })\mu .
\end{eqnarray}%
Here\ $\mu _{ii}(i=1,2,3)$ and $\mu _{ij}(i,j=1,2,3)$ are the relevant
elements of the Dirac- and Majorana-type magnetic moment matrices,
respectively \footnote{%
Ref. \cite{brxII} mistakenly put a factor of $P_{e1}^{2\nu }$ in the first
term of eq. (9).}.

Similarly, as developed in ref. \cite{pvogel} and applied in the refs. \cite%
{oleg,brxm}, the effective magnetic moment for the LMA-MSW solution in terms
of the flavor-basis can be expressed as%
\begin{equation}
(\mu _{eff}^{F})^{2}=\langle P^{m}\rangle _{ee}\mu _{e}^{2}+(1-\langle
P^{m}\rangle _{ee})(\cos ^{2}\theta _{23}\mu _{\mu }^{2}+\sin ^{2}\theta
_{23}\mu _{\tau }^{2})
\end{equation}%
where $\langle P^{m}\rangle _{ee}\ $is given in Eq. (7) and $\mu _{e},$ $\mu
_{\mu },\mu _{\tau }$ are the magnetic moments in the flavor bases.

\begin{table*}[t]
\begin{center}
\begin{tabular}{l|lll||lll||lll}
\hline \hline
Flux & $\left \vert \mu _{11}\right \vert $ & $\left \vert \mu _{22}\right
\vert $ & $\left \vert \mu _{33}\right \vert $ & $\left \vert \mu
_{12}\right \vert $ & $\left \vert \mu _{13}\right \vert $ & $\left \vert
\mu _{23}\right \vert $ & $\left \vert \mu _{e}\right \vert $ & $\left \vert
\mu _{\mu }\right \vert $ & $\left \vert \mu _{\tau }\right \vert $ \\ \hline
pp & $1.05$ & $1.58$ & $5.94$ & $8.60$ & $1.04$ & $1.52$ & $1.17$ & $1.73$ & 
$1.96$ \\ 
$^{7}$Be & $1.09$ & $1.47$ & $5.94$ & $8.60$ & $1.07$ & $1.43$ & $1.18$ & $%
1.70$ & $1.93$ \\ 
pep & $1.10$ & $1.44$ & $5.94$ & $8.60$ & $1.08$ & $1.40$ & $1.19$ & $1.69$
& $1.91$ \\ \hline \hline
\end{tabular}%
\end{center}
\caption{In the mass-basis, Dirac-type column (1-3), Majorana-type column
(4-6) and in the flavor basis (7-9) magnetic moments in units of $\  \times
10^{-11}\protect \mu _{B}$ derived from effective magnetic moment as shown in
Fig. I, for \textit{pp}, $^{7}$\textit{Be }and \textit{pep} of Borexino
phase-I and phase-II runs.}
\end{table*}

\section{Analysis and Results}

\textit{\ }We follow the BOREXINO's paper of phase-I \cite{brxpp,brxbe,brxpe}
and phase-II \cite{brxII} and take the number of \ target electrons per 100
tons, $N_{e}$ = $3.307\times 10^{31}$ while take the $pp$ reaction flux from
ref. \cite{bks}\  \ that is also summarized in the Appendix of our recent
work \cite{ANKS}. Since \ the $^{7}$\textit{Be} and $pep$ fluxes have a
discrete spectra, therefore we treat them as delta functions in our analysis
to evaluate the rate in Eq. (\ref{eq:totrate}). Following Borexino analysis,
we use the high-metallicity SSM flux values $\phi _{^{7}Be}=4.48\times
10^{9} $ cm$^{-2}$s$^{-1}$ at $0.862$ MeV and $\phi _{pep}=1.44\times 10^{8}$%
cm$^{-2}$s$^{-1}$ at 1.44 MeV in our calculations. For predicted rates for
each spectra, we use Eq. (7) and (8) to find the modifications due to the
small matter effects to the energy- independent vacuum value of $\langle
P^{vac}\rangle _{ee}$ = 0.558 and obtain the values $\langle P^{pp}\rangle
_{ee}$ = 0.554$\ $for \textit{pp}, $\langle P^{^{7}Be}\rangle _{ee}$= 0.536
for $^{7}$\textit{Be} and $\langle P^{pep}\rangle _{ee}$= 0.529$\ $for 
\textit{pep}. We can write down the basic structure of the expected rates to
compare them with Borexino's results as

\begin{equation}
R_{\nu }^{i}=N_{e}\int_{0}^{E_{max}}dE_{\nu }\phi ^{i}(E_{\nu })\left(
\sigma _{e}(E_{\nu })\langle P^{mi}\rangle _{ee}+\sigma _{\mu ,\tau }(E_{\nu
})[1-\langle P^{mi}\rangle _{ee}]\right) ,  \label{eq:totrate}
\end{equation}%
where $\langle P^{mi}\rangle _{ee}$ are given in eq. (7) and (8), with the
index \textit{i} indicating whether vac, \textit{pp}, $^{7}$Be or \textit{pep%
}. The cross-sections $\sigma _{e}(E_{\nu })$ and $\sigma _{\mu ,\tau
}(E_{\nu })$ are defined in Eqs. (1), (2) and (3) . To find the best-fits
values and limits on the parameters $\overrightarrow{\lambda }\equiv (\sin
^{2}\theta _{W},\mu _{\nu }^{eff},\left \langle r_{\nu }^{2}\right \rangle )$
, we use a $\chi ^{2}-$ estimator 
\begin{equation}
\chi ^{2}(\overrightarrow{\lambda })=\sum_{i}\frac{%
(R_{exp}^{i}-R_{pre}^{i}(1+\alpha ^{i}))^{2}}{(\sigma _{stat}^{i})^{2}}%
+\left( \frac{\alpha ^{i}}{\sigma _{\alpha }^{i}}\right) ^{2},
\label{eq:chi_ssq}
\end{equation}%
where $i$ runs over the solar neutrino sources \textit{pp}, $^{7}$\textit{Be}
and \textit{pep}. In eq. (15), $R_{exp}$\ are the experimental event rates
observed at Borexino in phase I and phase II with $\sigma _{stat}$\ as the
statistical uncertainty for each of the five experiments while $R_{pre}$\ is
the predicted event rate corresponding to each experiment calcuated from eq.
(14).\textbf{\ }With $\mu _{\nu }^{eff}=0,\left \langle r_{\nu }^{2}\right
\rangle =0$ and the PDG(2016) value $\sin ^{2}\theta _{W}$ = 0.2313 \cite%
{PDG16} , our predicted event rate values are $R^{pp}$ = 131$\pm $2.4 (144$%
\pm $ 13, 134$\pm $10), $R^{^{7}Be}$ = 47.8$\pm $2.9 (46$\pm $1.5, 48.3$\pm $%
1.1) and $R^{pep}$ = 2.74$\pm $0.05 (3.1$\pm $0.6, 2.43$\pm $0.36), where
the Borexino's phase-I and phase-II measured values are given inside the
parentheses, respectively. The measured values for phase-I were taken from
ref. \cite{brxpp,brxbe,brxpe}\ and for phase-II were taken from ref. \cite%
{brxII}. Notice that we have taken the uncertainties of the expected event
rates $2.4,2.9$ and $0.05$ for pp, $^{7}$Be and pep, respectively, from the
TABLE I of ref.\  \cite{brxII}. In Eq. (\ref{eq:chi_ssq}),\ we add penalty
term corresponding to each experiment to account for the theoretical
uncertainties coming from the solar flux models and from the oscillation
parameters.

We fit $\sin ^{2}\theta _{W},\mu _{\nu }^{eff}$ and $\left \langle r_{\nu
}^{2}\right \rangle $ in one and two parameter spaces while minimizing over
the pull parameters "$\alpha "$\ within$\  \sigma _{\alpha }^{i}$\ where $%
\sigma _{\alpha }^{i}$\ are the relative uncertainties calculated from the
total uncertainties $(2.4,2.9$ and $0.05)$ of the predicted event \textbf{%
rates} given in TABLE I of ref.\  \cite{brxII}, and also reported above in
addition to our predicted event rates. The calculated relative uncertainties
used in this analysis are 2\%, 6\%, 2\% for pp, $^{7}$Be and pep fluxes,
respectively. We have also included the radiative correction (2\%) to the SM
cross-section given in Eq. (2). Values of the mixing parameters were taken
from PDG(2016) \cite{PDG16}

Using Borexino's published values for the rates in phase-I \cite%
{brxpp,brxbe,brxpe} and phase-II \cite{brxII} and their 1$\sigma $
statistical uncertainties for $pp$, $^{7}$Be and $pep$ real time detections,
we find a best-fit value $\sin ^{2}\theta _{W}=$0.235$\pm $0.019, which is
consistent with both the low-energy theoretical prediction \cite{erler} and
with $\overline{MS}$ value at the Z-boson mass.\textbf{\ }This result has a
slightly weaker precision than those obtained from decay and reactor data
studies \cite{porsev,amirgl, deniz2,vallewm}. Because it includes the $pp$\
and $^{7}Be$\ data, our determination of \ $\sin ^{2}\theta _{W}=$0.235$\pm $%
0.01 is below the energies of its other measurements to date, excepting the
recent estimate in ref. \cite{ANKS} for phase-II of Borexino's data.
Previously, the lowest energy determinations were the atomic parity
violation measurement in $^{133}$Cs at 2.4 MeV \cite{Cs,boulder,porsev}.\ 

In Fig. 1, we show the 2-\textit{d.o.f }parameter space of $\sin ^{2}\theta
_{W}$ and the neutrino effective magnetic moment $\mu _{\nu }^{eff}$ with
the 90\%, 95\% and 99\% C.L. boundaries from inner to outer, respectively.
In Fig. 2, we demonstrate the 1-dim $\Delta \chi ^{2}$ distribution of the
three unknown parameters, $\sin ^{2}\theta _{W}\ ($top$),\mu _{\nu }^{eff}($%
middle$),\left \langle r_{\nu }^{2}\right \rangle $ (bottom) with 1$\sigma $
and the 90\% C.L. projection on each distribution. The best-fits and bounds
of the three parameters at 90\% C.L. are $\sin ^{2}\theta _{W}=$0.235$\pm $%
0.019, $\mu _{\nu }^{eff}\leq 8.7\times 10^{-12}\mu _{B}\ $and$\ -0.82\times
10^{-32}$cm$^{2}\leq \left \langle r_{\nu _{e}}^{2}\right \rangle \  \leq
1.27\times 10^{-32}$ cm$^{2}\ $and\ $-9\times 10^{-32}$cm$^{2}\leq \left
\langle r_{\nu _{\mu },\nu _{\tau }}^{2}\right \rangle \  \leq 3.1\times
10^{-31}$ cm$^{2}$. The two parameter fits in Fig. 1 and one parameter fits
in Fig. 2 for $\sin ^{2}\theta _{W}$, $\mu _{\beta }^{eff}$ and for $\left
\langle r_{\nu _{e}}^{2}\right \rangle $ and $\left \langle r_{\nu _{\mu
}}^{2}\right \rangle $ or $\left \langle r_{\nu _{\tau }}^{2}\right \rangle $
are consistent with each other. The best-fit parameter values and bounds are
also quoted in TABLE I\textbf{.} Notice that we have also carried out an
analysis for the phase-II\ data only and found that upper limit value of $%
\mu _{\nu }^{eff}$\ obtained is $1.3\times 10^{-11}\mu _{B}$ which is around
a factor of 2 different than the value quoted in ref. \cite{brxm} by
Borexino collaboration. The apparent difference occurs due to the fact that
the Borexino collaboration includes both statistical and systematic
uncertainties in their analysis while we use only the statistical
uncertainties. This could be a good basis of motivation for the Borexino
experiment to perform a detailed spectral analysis using the combined data
of phase-I and phase-II.

The upper bound value of $\mu _{\nu }^{eff}$ at 90\% C.L. is then used in
Eq. (11) and (12) to derive the values of the Dirac-type and Majorana-type
magnetic moment matrix elements by taking one element at-a-time, the
relevant parameter values obtained are quoted in TABLE II. Since we have
three different solar fluxes with different energies so we derive the matrix
element values for each flux separately and marginal differences from lower
to higher energy fluxes can be seen from the Table. \textbf{A }similar
procedure has been repeated for the neutrino magnetic moment in the flavor
basis using Eq. (13) and the values obtained are given in the last three
columns of TABLE II.

\section{Conclusions}

\textit{\ }In this paper we have performed event rate analysis to estimate
the value of electroweak parameter $\sin ^{2}\theta _{W}$ and
electromagnetic parameters of neutrino due to its nonzero size using the
lowest available solar neutrino energy spectra of\textit{\ pp, }$^{7}$%
\textit{Be}~and \textit{pep} from Borexino phase I and phase II runs. We
have included both the theoretical model and the measured statistical
uncertainties to estimate the best-fit values and bounds of the parameters.
The best-fit value of $\sin ^{2}\theta _{W}$\ obtained at 90\% C.L. was 0.235%
$\pm $0.019 with the precision better than 8\%. The precision obtained here
is comparable to that obtained from the combined reactor and accelerator
very short-baseline experiments \cite{amirgl}. The upper bound obtained for
the effective magnetic moment parameter is $\mu _{\nu }^{eff}\leq 8.7\times
10^{-12}\mu _{B}$\ at 90\% C.L., which has a factor of 3 improvement
compared to the previous bounds. It turns out in improving the bounds of the
magnetic moment elements for both the Dirac-type and Majorana-type in the
mass and also in the flavor basis by a factor of 3 to 5 as given in TABLE
II. Similarly, the bounds on the neutrino charge radii turn out to be$\
-0.82\times 10^{-32}$cm$^{2}\leq \left \langle r_{\nu _{e}}^{2}\right
\rangle \  \leq 1.27\times 10^{-32}$ cm$^{2}\ $and\ $-9\times 10^{-32}$cm$%
^{2}\leq \left \langle r_{\nu _{\mu },\nu _{\tau }}^{2}\right \rangle \  \leq
3.1\times 10^{-31}$ cm$^{2}\ $at 90\% C.L..

Current and future solar neutrino low-energy fluxes, \textit{pp }and~$^{7}$%
\textit{Be} in particular, have a strong potential to provide a precise
measurement of $\sin ^{2}\theta _{W}$ and higher sensitivity to all of the
electromagnetic properties of neutrinos and the related parameters to give
decisive information about the nature of neutrinos whether Dirac or
Majorana. They can test sin$^{2}\theta _{W}$ = 0.23867$\pm $0.00016
prediction of $\overline{MS}$ running of the parameter to the sub-MeV
region. Moreover, from the low-energy solar fluxes the electromagnetic
parameters of at least two flavors of neutrinos can be constrained and
explored as we showed in this work. We anticipate that the Borexino
collaboration will carry out a detailed spectral analysis for the combined
data of phase I and phase II to confirm the calculations of this work.

\begin{acknowledgments}
The author thanks Prof. Douglas McKay from Kansas University and\textbf{\ }%
Prof. Jiajae Ling from Sun Yat-Sen University for their useful suggestions
and comments. The financial support for this work has been provided by the
Sun Yat-Sen University under the Postdoctoral Fellowship program, the China
Postdoctoral Science Foundation under the grant \# 74130-41090002 and the
NPC fellowship award, Neutrino Physics Center, Fermilab.
\end{acknowledgments}

\end{document}